\documentclass[aps,pra, twocolumn,nofootinbib, showpacs]{revtex4-1} 
\usepackage[unicode=true,pdfusetitle, bookmarks=true,bookmarksnumbered=false,bookmarksopen=false, breaklinks=false,pdfborder={0 0 0},backref=false,colorlinks=false] {hyperref}
\hypersetup{ colorlinks,linkcolor=myurlcolor,citecolor=myurlcolor,urlcolor=myurlcolor}
\usepackage{braket,colortbl,amsthm,amsmath,cleveref,amssymb,txfonts}
\definecolor{myurlcolor}{rgb}{0,0,0.7}
\usepackage{graphics,graphicx}
\usepackage{color}
\usepackage{graphicx}
\usepackage[format = plain,labelfont = bf,up, textfont = normal , up, justification =raggedright, singlelinecheck =false]{caption}
\usepackage{subcaption}
\usepackage{tabularx}
\usepackage{amsmath}

\usepackage{graphicx}
\usepackage[utf8x]{inputenc}
\usepackage{color}
\usepackage{braket}
\usepackage{latexsym}
\usepackage{amssymb}
\usepackage{amsthm}
\usepackage{bm}
\usepackage{graphics,epstopdf}
\usepackage{color}
\usepackage{enumitem}
\usepackage{fmtcount}
\usepackage{booktabs}

\usepackage{epsfig}

\theoremstyle{plain}

\def\bea{\begin{eqnarray}}
\def\eea{\end{eqnarray}}
\def\ba{\begin{array}}
\def\ea{\end{array}}

\def\ket{\rangle}
\def\bra{\langle}
\def\beq{\begin{equation}}
\def\eeq{\end{equation}}

\begin{document}

\title{Sequential measurement-device-independent entanglement detection by multiple observers}

\author{Chirag Srivastava, Shiladitya Mal, Aditi Sen(De), and Ujjwal Sen}

\affiliation{Harish-Chandra Research Institute, HBNI, Chhatnag Road, Jhunsi, Allahabad 211 019, India}

\begin{abstract}

Violation of a Bell inequality certifies that the underlying state must be entangled in a device-independent way, although there may exist some entangled states which do not violate such an inequality. On the other hand, for every entangled state, it is possible to find a hermitian operator called entanglement witness that can detect entanglement through some local measurements in a device-dependent method.  The methods are significantly fragile to lossy detectors. To avoid such difficulties, measurement-device-independent entanglement witness based on a semi-quantum nonlocal game was proposed which turns out to be robust against lossy detectors. We employ here such a measurement-device-independent entanglement witness to detect entanglement in a scenario where half of an entangled pair is possessed by a single observer while the other half is with multiple observers performing measurements sequentially, independently, and preserving entanglement as much as possible. Interestingly, we find that the numbers of successful observers who can detect entanglement measurement device-independently, both with equal and unequal sharpness quotients, are higher than that obtained  with standard and Bell inequality-based entanglement detection methods,  reflecting its robustness. The entanglement contents of the sequentially shared states are also analyzed. Unlike other scenarios, our investigations also reveal that in this measurement-device-independent situation, states having entanglement in proximity to maximal, remains entangled until two sequential observers even if they measure sharply.
\end{abstract}


\maketitle
\section{Introduction}
The existence of entangled states \cite{Horodecki'09} is one of the most nonclassical features of the quantum mechanical description of nature, which, e.g., can lead to violation of Bell inequality \cite{Bell'64}, testable in the laboratory. The violation implies that quantum theory can not be replaced by a local realistic model, compatible with classical theory, and this feature of quantum theory enables various quantum information processing tasks like generation of true randomness \cite{rand'10}, secure key distribution \cite{BB84}, and possibly also related to quantum communication \cite{Tele'93}. 

Over the years, it has been established that one of the efficient ways to detect entanglement in the laboratory is through entanglement witnesses (EWs), which can be implemented through local measurements performed on the individual systems constituting the composite system \cite{Horodecki'96, Terhal}. However, implementing an EW requires proper characterization of the measurement devices and some prior information of the shared states. On the other hand, violation of Bell inequality certifies entanglement device-independently while paying a cost, viz., there exist entangled states, where entanglement, seemingly, cannot be probed via violation of a Bell inequality \cite{Werner'89}. See \cite{Methot'07} in this regard.

It is known that corresponding to every Bell inequality, there is a nonlocal game, and for each nonlocal game, one can construct a Bell inequality \cite{Silman'08}. Extending the Bell scenario, Buscemi has recently proposed a ``nonlocal semi-quantum game'' where every entangled state yields a higher pay-off compared to all separable states \cite{Buscemi'12}. In this game, two observers share a bipartite state and on top of that, instead of classical inputs, like in a standard Bell scenario, a ``referee'' gives them quantum inputs. Each party then measures jointly on the respective inputs and their part of the shared state. Outputs of the observers together with inputs are used to constitute the pay-off function which shows advantage for any entangled state over all separable states. Note that except for the quantum inputs, other devices are untrusted in this scenario. Such a semi-quantum nonlocal game can also be extended to the multiparty scenario \cite{Buscemi'12}.

Since all entangled states are ``nonlocal'' according to the semi-quantum game \cite{Buscemi'12}, i.e., since the pay-off function provides a higher value for any entangled state than all separable states, it can be a witness for detecting entanglement. We refer to such a situation -- a higher value of the pay-off function than all separable states -- as ``Buscemi-nonlocality'', and is to be contrasted with the previous notion of ``Bell-nonlocality'' \cite{Bell'64}, which referred to a violation of Bell inequality. It is known that there exist EWs for every entangled state \cite{Horodecki'96,Terhal,Guhne'03}. Given such an EW, in \cite{Cyril'13}, Branciard \emph{et al.} constructed a new EW, based on Buscemi's game, which does not depend on the internal functioning of measurement devices. Along with the inconveniences mentioned before, detecting entanglement through standard EWs or Bell inequality violation has another drawback, viz., that they are not robust against imperfections in detectors. Specifically, lossy detectors can wrongly indicate a separable state as entangled \cite{Bruss'07,Kornikar}. Measurement-device-independent EWs (MDI-EWs) never announce separable states as entangled \cite{Cyril'13}.    

In quantum information processing tasks, it is important to distribute resource states among several parties. In the literature, there are various protocols to do that. In \cite{silva}, Silva \emph{et al.} showed that when an entangled pair is shared between a single observer (say, Alice) at one side and several other observers (say, Bobs) at the other, acting sequentially and unsharply, no more than two Bobs can exhibit violation of the Bell-CHSH inequality with Alice \cite{CHSH}. See also \cite{mal}, and for experimental verification, see \cite{Ex1, Ex2}. Later, the concept of sequential unsharp action has been extended to other contexts, like Bell-type inequalities with more than two settings at each site \cite{Das'19}, quantum steering \cite{Sasmal'18, Shenoy'19} and entanglement witnesses \cite{bera}. Recently, it has also been applied in the scenarios of random access codes \cite{Mohan'19} and in re-usability of teleportation channels \cite{Roy'19}.

In the present work, we investigate how the power of MDI-EWs can be reflected in a resource distribution protocol. In \cite{bera}, it was found that at most twelve Bobs can detect entanglement sequentially, when standard EWs were employed. We also consider pure entangled state as the inital state and find the maximal number of Bobs allowed in this protocol. The behaviour of entanglement content of the subsequent shared states is also observed. We find that the maximal number of Bobs who can identify entanglement with a single Alice can go upto fourteen in a measurement-device-independent way when the shared initial state has entanglement more than or equal to 93.5\% of the singlet. We also study the case when all the Bobs measure with a common sharpness parameter. For an initially shared maximally entangled state, the  maximum number of Bobs who can sequentially detect entanglement while using a common sharpness parameter is six, which is greater than when the same task is considered with standard EWs.  


 It is to be mentioned here that in the case of detecting entanglement, using unsharp versions of EWs \cite{bera}, if any of the Bobs measures sharply, i.e., projectively, then there is no possibility of detecting entanglement by any subsequent Bob, as there is no residual entanglement between Alice and the subsequent Bob. Therefore, if the Bobs have to detect entanglement sequentially, then all of them except the last one must measure unsharply \cite{Busch'96}. On the other hand, at each step, a very unsharp measurement may rule out the possibility of detecting entanglement, and this can be interpreted as another face of the well-known trade-off between information gain and disturbance \cite{Fuchs'96}. Hence, every Bob has to measure with a \emph{threshold} sharpness parameter, so that Alice and he can detect entanglement in the sequential process. Interestingly, in the context of MDI-EWs, we find that even if the first Bob measures sharply, then the second Bob also can detect entanglement, which was not the case when standard EWs were employed \cite{bera}. 

This paper is organised as follows. In Sec. \ref{bristi}, we briefly discuss MDI-EWs and the unsharp measurement formalism adopted for the purpose of our work. In Sec. \ref{aandhi}, the scenario of entanglement sharing in the context of MDI-EW is  discussed. In Sec. \ref{optimal_case}, we find  the maximum number of sequential and independent single-lab observers, who are able to detect the bipartite entanglement shared with the common distant-lab observer using MDI-EWs.
In Sec. \ref{alternate}, we analyze the change in entanglement content due to an unsharp measurement required for the MDI-EW procedure, in the states shared between the common observer and the sequential observers.
In Sec. \ref{baadal}, the case of sequential observers measuring with equal sharpness is considered, and finally we end with conclusion in Sec. \ref{tareef}.

\section{Essentials}
\label{bristi}
Let us begin by discussing the necessary ingredients required to detect bipartite entanglement shared between Alice at the one side and multiple Bobs at the other side in a measurement-device-independent scenario. 
\subsection{Measurement-device-independent entanglement witness}
\label{MDI-EW}
  An entanglement witness operator, $W$, is defined as a hermitian operator such that for all states $\sigma_{AB}\in\mathcal{S}$, tr$(\sigma_{AB}W)\geqslant 0$, while there exists at least one entangled state, $\rho_{AB}$, in the same bipartition, such that tr$(\rho_{AB}W)<0$, where $\mathcal{S}$ is the set of separable states in the bipartition, $A:B$  \cite{Horodecki'96,Terhal,Guhne'03}. But such witness operators have at least two disadvantages. Firstly, to implement them, one requires characterized devices as well as some prior information about the state to be detected, and secondly, in the case of lossy measurements, the expectation value of witness operators for separable states may turn out to be negative, leading to a false positive detection of entanglement \cite{Bruss'07}. 
 
 To avoid such uncertainties, Branciard $et~al.$ introduced the concept of measurement-device-independent entanglement witnesses \cite{Cyril'13}. Specifically, given an EW, the semi-quantum nonlocal game of Buscemi \cite{Buscemi'12} is used to obtain as MDI-EW. Consider the scenario where two parties, Alice and Bob, possess a shared entangled state $\rho_{AB}$ operating on the Hilbert space $\mathcal{H}_A\otimes\mathcal{H}_B$, with the dimensions of $\mathcal{H}_A$ and $\mathcal{H}_B$ being 2 each. Further, Alice and Bob receive quantum inputs, from a ``referee'', in the form of a state from a set of qubit states denoted as $\{\tau_s\}_{s=0}^{3}$ and $\{\omega_t\}_{t=0}^{3}$ respectively. They then perform a joint measurement on their respective parts of the shared state and the state obtained from the referee, with the referee providing the state randomly from the respective sets. The conditional probability that Alice and Bob obtain the classical outcomes $a$ and $b$ respectively, given that the input states to them are respectively $\tau_s$ and $\omega_t$, is denoted by $P(a,b|\tau_s,\omega_t)$. 

Let us now consider a situation where one chooses joint measurements that have only two outcomes, viz., i.e. either 0 or 1.
The ``MDI-EW function'' for the state $\rho_{AB}$ is then shown to be given by \cite{Cyril'13}
\begin{equation}
\label{genI}
I(\rho_{AB})=\sum_{s,t} \beta_{st}P(1,1|\tau_s,\omega_t).
\end{equation}
Here,
\begin{equation}
\label{WS_CP}
P(1,1|\tau_s,\omega_t)=\text{tr}[(|\Phi^+\ket\bra\Phi^+|\otimes|\Phi^+\ket\bra\Phi^+|)(\tau_s\otimes\rho^w_{AB}\otimes\omega_t)],
\end{equation}
where outcome 1 indicates the successful projection of the joint measurements by any observer on her or his respective part of the shared state and an input state onto the maximally entangled state,  $|\Phi^+\ket=\frac{1}{\sqrt{2}}(|00\ket+|11\ket)$. The real coefficients, $\beta_{st}$, are set by standard EW.
Like the EW operator $W$, $I\geqslant 0$ for any separable state and negative for at least one entangled state. It was shown that the MDI-EW can easily be generalized to higher dimensions and to cases of states with a higher number of parties \cite{Cyril'13}.

\subsection{Unsharp measurement and modified MDI-EW}

It can be seen that evaluation of the MDI-EW function, $I(\rho_{AB})$, given by Eq. (\ref{genI}), requires a two-outcome projective measurement, with projectors 
\begin{eqnarray}
\label{sm}
 \mathcal{P}^+=|\Phi^+\ket\bra\Phi^+|, \nonumber \\
 \mathcal{P}^-=\mathbb{I}_4-|\Phi^+\ket\bra\Phi^+|, 
 \end{eqnarray}
 where $\mathcal{P}^+$ and $\mathcal{P}^-$ are assumed to correspond to outcomes 1 and 0 respectively. 
 Let us now consider an unsharp version of the above projective measurement, described by ``effect'' operators $\{\mathcal{E}^+_\lambda,~\mathcal{E}^-_\lambda\}$, relative to $\{\mathcal{P}^+,~\mathcal{P}^-\}$, given by
 \begin{eqnarray}
 \label{wm}
 \mathcal{E}^+_\lambda&=&\lambda\mathcal{P}^+ +\frac{1-\lambda}{4}\mathbb{I}_4, \nonumber \\ 
 \mathcal{E}^-_\lambda&=&-\lambda\mathcal{P}^+ +\frac{3+\lambda}{4}\mathbb{I}_4
 \end{eqnarray}
 where $\mathcal{E}^+_\lambda$ and $\mathcal{E}^-_\lambda$ correspond to outcomes 1 and 0 respectively, and  $0\leqslant \lambda\leqslant 1.$ $\mathbb{I}_d$ denotes the identity operator on $\mathbb{C}^d$.
 Given the situation that Alice and Bob respectively receive states $\tau_s$ and $\omega_t$ as inputs and Alice measures in $\{\mathcal{P}^+,~\mathcal{P}^-\}$ on her part of $\rho_{AB}$ and $\tau_s$ while Bob performs an unsharp measurement with $\{\mathcal{E}^+_\lambda,~\mathcal{E}^-_\lambda\}$ on his part of $\rho_{AB}$ and $\omega_t$, the conditional probability that Alice and Bob both obtain outcome 1 is then given by
\begin{equation}
\label{wCP}
 P_\lambda(1,1|\tau_s,\omega_t)=\text{tr}[(\mathcal{P}^+\otimes\mathcal{E}^+_\lambda)(\tau_s\otimes\rho_{AB}\otimes\omega_t)].
\end{equation}
Therefore, the modified MDI-EW function for the case when one of the parties perform an unsharp measurement, with $\{\mathcal{E}^+_\lambda,~\mathcal{E}^-_\lambda\}$, on his part of the state $\rho_{AB}$ and the input from the referee, reads
\begin{equation}
\label{wgenI}
I_\lambda(\rho_{AB})=\sum_{s,t} \beta_{st}P_\lambda (1,1|\tau_s,\omega_t).
\end{equation}

{\textbf{Post-measurement state}:} In our sequential-measurement scenario, the post-measurement state plays an important role and hence let us identify the rule for assigning the post-measuement state to a measurement outcome. Suppose an unsharp joint measurement with $\{\mathcal{E}^+_\lambda,~\mathcal{E}^-_\lambda\}$ is performed at one side of the shared state and on the quantum input, denoted by $\eta$. According to the von Neumann-L\"uders transformation rule \cite{Busch'86}, up to a unitary, if the $'+'$ outcome occurs, the post-measurement state is given by 
\begin{equation}
\label{genpmS}
\left(\mathbb{I}\otimes\sqrt{\mathcal{E}^+_\lambda}\right)~\eta\left(\mathbb{I}\otimes\sqrt{\mathcal{E}^+_\lambda}\right).
\end{equation}

\subsection{MDI-EW for Werner states}
Consider the (bipartite) Werner states, given by
\begin{equation}
\label{WS}
\rho^w_{AB}=q~|\Psi^{-}\ket\bra\Psi^{-}|+\frac{1-q}{4}\mathbb{I}_4,   
\end{equation}
where $q$ is the mixing probability of the singlet, $|\Psi^-\ket=\frac{1}{\sqrt{2}}(|01\ket-|10\ket)$, in $\rho^w_{AB}$.
The MDI-EW function, $I(\rho^w_{AB})$, for this state can be represented as \cite{Cyril'13} 
\begin{equation}
\label{WS_I}
I(\rho^w_{AB})=\frac{5}{8}\sum_{s=t}P(1,1|\tau_s,\omega_t)-\frac{1}{8}\sum_{s\neq t}P(1,1|\tau_s,\omega_t),
\end{equation} 
where $s,~t$ take values 0, 1, 2, and 3, and
\begin{equation}
\label{tau}
\tau_s=\sigma_s\frac{\mathbb{I}_2+\vec{\sigma}\cdot \vec{n}}{2}\sigma_s,~~~\omega_t=\sigma_t\frac{\mathbb{I}_2+\vec{\sigma}\cdot \vec{n}}{2}\sigma_t,
\end{equation}
with $\sigma_0=\mathbb{I}_2$,  $\vec{\sigma}=(\sigma_1,\sigma_2,\sigma_3)$ being the usual Pauli matrices, and $\vec{n}=\frac{1}{\sqrt{3}}(1,1,1)$.
 
 The expression for the MDI-EW function in Eq. (\ref{WS_I}) can be simplified and written in terms of the state parameter $q$, as  
\begin{equation}
\label{WS_I2}
I(\rho^w_{AB})=\frac{1-3q}{16}.
\end{equation}
This will be useful in our later calculations.

\section{Scenario}
\label{aandhi}

 
Consider a scenario where, initially, a bipartite entangled state is shared between two spatially separated laboratories, overseen respectively by Alice ($A$) and the Bobs $(B_i,~i=1,2,\ldots,n)$. $A$ measures projectively on her part and several Bobs ($B_i$), in the other laboratory, measure sequentially and independently. The aim is to find the $maximum$ number of Bobs, $n$, such that each $AB_i$ pair is able to witness Buscemi-nonlocality or MDI entanglement between them.
As operations are local and strong enough to fetch information about the entanglement content of a state shared by $A$ and $B_i$, it is expected that the state shared by $A$ and $B_{i+1}$ (next Bob in sequence) will have less entanglement than that of $AB_i$. The unsharp measurement has to be strong enough to detect the shared state's entanglement, and at the same time, it has to be weak enough so that the post-measured state shared between Alice and the next Bob retains as much entanglement as possible, so that the remnant resource can be used subsequently. 
This observation tells us that there may exist an upper bound on the maximum number of Bobs, such that each of them can detect entanglement by combining their and Alice's statistics. Note that $A$ can do her part of measurements at any time, i.e. independent of any of the $B_i$'s measurement, as operators from $\mathcal{H}_A$ and those from $\mathcal{H}_B$ commute with each other.  

 

\subsection{Subsequent shared states due to unsharp measurement}

Let us consider the cases where Alice, $A$, and the first Bob, $B_1$, share a pure entangled state. It will be seen in the following paragraph that subsequent weak measurements by each observer produces a mixed state with a mixture of initial entangled state, shared by the $AB_1$ pair, and  white noise. 

Suppose that Alice and the first Bob share the state, $|\Psi\ket=\alpha|01\ket-\sqrt{1-\alpha^2}|10\ket$, for $0 < \alpha \leqslant \frac{1}{\sqrt{2}}$, or equivalently, $\rho^{w_\alpha}_{AB_1}$, being given by
\begin{equation}
\label{WS2}
 \rho^{w_\alpha}_{AB_1}=q_1|\Psi\ket\bra\Psi|+\frac{1-q_1}{4}\mathbb{I}_4, 
\end{equation} 
 with $q_1=1$. Now, let $B_1$ measure the unsharp POVM with effect operators $\{\mathcal{E}^+_{\lambda_1},~\mathcal{E}^+_{\lambda_1}\}_{BB'}$, and sharpness parameter $\lambda_1$, on his part of the shared state and input system $B'$ (from the referee) in state $\omega_t$. The quantum input $\omega_t$ has four random choices, say, for each $t=0,~1,~2,$ and 3, occuring with equal probability. As each Bob measures independently, the average state $\rho_{AB_2}$ that $A$ and the next Bob $B_2$ share, is given by
\begin{multline}
\text{tr}_{B'}\bigg\{\frac{1}{4}\sum_{t=0}^{3}\Big[(\mathbb{I}_2
\otimes\sqrt{\mathcal{E}^+_{\lambda_1}})(\rho^{w_\alpha}_{AB_1}\otimes\omega_t)(\mathbb{I}_2
\otimes\sqrt{\mathcal{E}^+_{\lambda_1}})\\
  +  (\mathbb{I}_2
\otimes\sqrt{\mathcal{E}^-_{\lambda_1}})(\rho^{w_\alpha}_{AB_1}\otimes\omega_t)(\mathbb{I}_2
\otimes\sqrt{\mathcal{E}^-_{\lambda_1}}) \Big]\bigg\},
\end{multline}
simplifying which, again turns out to be a state of form in Eq. (\ref{WS2}), viz., 
$$\rho^{w_\alpha}_{AB_2}=q_2|\Psi\ket\bra\Psi|+\frac{1-q_2}{4}\mathbb{I}_4,$$ 
with
$q_2=f(\lambda_1)q_1,$ where
\begin{equation}
\label{fof_lam}
f(\lambda)=\frac{1}{2}\bigg[1+\frac{\sqrt{(1+3\lambda)(1-\lambda)}+\sqrt{(3-3\lambda)(3+\lambda)}}{4}\bigg].
\end{equation}

The above structure is iterative, and therefore, the state that the $AB_i$ duo possesses, reads
$$\rho^{w_\alpha}_{AB_i}=q_i|\Psi\ket\bra\Psi|+\frac{1-q_i}{4}\mathbb{I}_4,$$
where
\begin{equation}
\label{q_iter}
q_i=f(\lambda_{i-1})q_{i-1},
\end{equation}
with $f(\lambda_i)$ being given in Eq. (\ref{fof_lam}).


\subsection{ Modified MDI-EW for non-maximally entangled states mixed with white noise}
In this subsection, we show how MDI-EW can be modified for unsharp measurements on a shared state $\rho^{w_\alpha}_{AB}$.
Note that the Werner states, $\rho^w_{AB}$, are a mixture of a singlet with white noise. For this class, the MDI-EW, $I(\rho^w_{AB})$, is an optimal witness \cite{Cyril'13, sanpera}.
 
Further, as the MDI-EW is independent of measurements, the bound for separable states remains zero even when one of the parties perform unsharp measurements.  
Hence, the modified measurement-device-independent entanglement witness for states, $\rho^{w_\alpha}_{AB}$, with $B$ doing an unsharp measurement is given by
\begin{equation}
\label{wWS_I}
I_\lambda(\rho^{w_\alpha}_{AB})=\frac{5}{8}\sum_{s=t}P_\lambda(1,1|\tau_s,\omega_t)-\frac{1}{8}\sum_{s\neq t}P_\lambda(1,1|\tau_s,\omega_t),
\end{equation}
where $P(1,1|\tau_s,\omega_t)$ for state $\rho^w_{AB}$ in Eq. (\ref{WS_I}) is just replaced by $P_\lambda(1,1|\tau_s,\omega_t)$ for state $\rho^{w_\alpha}_{AB}$.
We find that
\begin{equation}
\label{wWS_I2}
I_\lambda(\rho^{w_\alpha}_{AB})=-\frac{\lambda q \alpha\sqrt{1-\alpha^2}}{4}+\frac{1-\lambda q}{16}.
\end{equation}
It can be seen that for $\lambda=1$ and $\alpha=\frac{1}{\sqrt{2}}$, Eq. (\ref{wWS_I2}) reduces to Eq. (\ref{WS_I2}).
It gives a lower bound on the sharpness parameter, which we refer to as the ``threshold sharpness parameter'', $\lambda^{th}=\frac{1}{q(1+4\alpha\sqrt{1-\alpha^2})}$, such that $I_\lambda(\rho^{w_\alpha}_{AB})<0$, $\forall \lambda>\lambda^{th}.$
Note that for the maximal resourceful state, i.e. the singlet, $\lambda^{th}=\frac{1}{3},$ which is the lowest for any entangled state of the form $\rho^{w_\alpha}_{AB}$.

\section{Witnessing Buscemi-nonlocality sequentially with initially shared entangled pure state}
\label{optimal_case}
We now move on to study the maximum number $(n)$ of Bobs who can act independently and sequentially to witness shared entanglement with a single observer, Alice, in the measurement-device-independent scenario. Note that this maximum is achieved only when all of the Bobs measure with their respective threshold sharpness parameters. The initial state shared between the two labs is assumed to be pure entangled state, $|\Psi\ket=\alpha|01\ket-\sqrt{1-\alpha^2}|10\ket$, where $0<\alpha \leqslant 1/\sqrt{2}$. The entanglement in this state is measured by the von Neumann entropy of the reduced density matrix (entanglement entropy) and is given by 
$E(\alpha)=-\alpha^2\log_2{\alpha^2}-(1-\alpha^2)\log_2{(1-\alpha^2)}.$   

\begin{figure}[h]
\includegraphics[width = 0.35\textwidth, angle=-90]{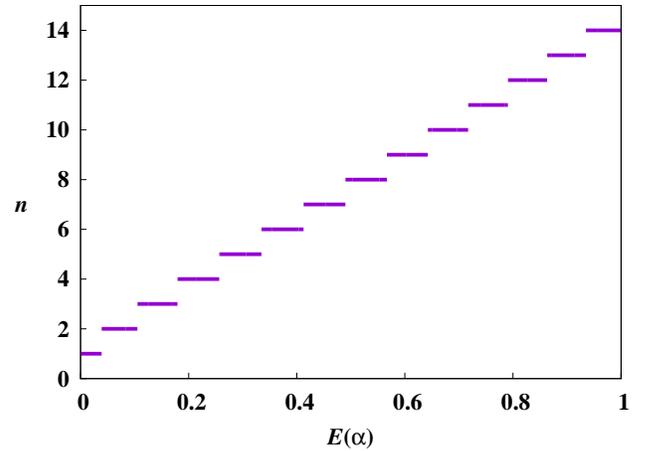} 
\caption{Sequential witnessing of entanglement in a measurement-device-independent scenario. We plot here the maximum number, $n$ in the MDI scenario vs. the entanglement, $E(\alpha)$, of the initial shared state $|\Psi\ket$. The vertical axis is dimensionless, while the horizontal one is in ebits.
}
\label{Fig.1}
\end{figure}

In Fig. \ref{Fig.1}, we depict the maximum number of Bobs that can detect entanglement in an MDI-way for a given entanglement content $E(\alpha)$. In particular, we find that if the initial shared state is close to the maximally entangled state, viz. if $E(\alpha)\gtrapprox 0.9349$, the maximum number of Bobs, $n$, which can keep the state entangled, reaches fourteen, the highest in the given scenario. In a similar study using standard EWs \cite{bera}, that can be termed as a ``device-dependent (DD) scenario'', the maximum number again remains fixed for a certain finite range of initial shared entanglement (of the $AB_1$ pair). However, it is interesting to notice that in the DD case, the maximum number of Bobs that can identify entanglement is twelve, which is less than that in the MDI scenario considered here. 
For any initially shared pure entangled state, the maximum number of Bobs in the device-dependent scenario of Ref. \cite{bera}, denoted by $n^{DD}$, is either less or equal to that in the MDI scenario considered here, i.e.,  $n^{DD}\leqslant n$ for any given initial entanglement. 

The lower value of \(n^{DD}\) than \(n\) for arbitrary initially shared pure entanglement deserves a comment. This is arguably due to the fact that the number of successful Bobs detecting entanglement with a single Alice depends on the choice of the witness, and in particular, on how the measurement disturbs the shared state. In the standard DD scenario considered in Ref. \cite{bera}, the sharp limit of the unsharp measurements are rank-one projective measurements, while the MDI scenario considered here involves quantum inputs, and the sharp measurement limit on the portion of the shared state in possession of the Bobs becomes a POVM of non-unit rank. A non-unit rank measurement has a general tendency of less affecting the entanglement of the shared state, and potentially affects the MDI procedure when we are far from the beginning Bob in the sequence of Bobs. It is to be remembered that the later Bobs are required to make sharper and sharper measurements to detect entanglement. Another point to mention in this respect is that the MDI scenario uses quantum inputs at both the labs possessing the bipartite state, and the subsequent measurements at both the labs are on \(\mathbb{C}^2 \otimes \mathbb{C}^2\). In contrast, the DD scenario of Ref. \cite{bera} considers single-qubit measurements at both the labs. Consequently, a comparison between the two scenarios is made difficult by another roadblock. 

 Comparing the results in \cite{bera} with ours, we can comment that the MDI scenario for witnessing entanglement is more robust in the context of unsharp measurements than that of the device-dependent witness. As we mentioned earlier MDIEW was also shown to be robust against standard EW in the case of lossy detectors \cite{Cyril'13}.

\section{Reduction in entanglement by unsharp measurement: Limit on successful detection of entanglement sequentially}
\label{alternate}  
In this section, we study the entanglement content of bipartite states shared by Alice and each of the sequential Bobs.
We investigate the reduction in bipartite entanglement, occuring due to the unsharp measurement performed at one side. 
For this purpose, we calculate the negativity \cite{negativity}, $N$, which for the state, $\rho^{w_\alpha}_{AB_i}$, is given by
\begin{equation}
 N(\rho^{w_\alpha}_{AB_i})=\max\left\{\frac{q_i(1+4\alpha\sqrt{1-\alpha^2})-1}{4},0\right\}.
\end{equation} 
For simplicity of notation, we will use $N_i$ instead of $N(\rho^{w_\alpha}_{AB_i})$, $i=1,2,\ldots,n$.
 The change in the negativity, denoted by $\Delta N_i(\lambda_i)$, due to an unsharp measurement by $B_i$ which is ``valid"  for $0\leqslant\lambda_i \leqslant 1$ and ``required'' to satisfy $\lambda_i > \lambda_i^{th}$ to witness entanglement in $\rho^{w_\alpha}_{AB_i}$, is defined as the difference in the negativities of the states before and after this measurement, i.e.,
\begin{equation}
\Delta N_i(\lambda_i)=N_i-N_{i+1}.
\label{delN}
\end{equation}
Here, by ``valid'', we mean that the parameters $\lambda_i$ in the measurement are to be chosen in the given range $(0\leqslant \lambda_i \leqslant 1)$ for the measurement to be to be quantum mechanically allowed, and by ``required'', we mean that the sharpness parameters $\lambda_i$ are to be chosen such that entanglement present can be detected. 
   Surely, $\Delta N_i(\lambda_i)$ is a positive quantity, as local measurements can only keep or decrease entanglement.
The negativity of the state that observers $A$ and $B_{i+1}$ share, can be obtained by the above equation, if one knows the negativity of the state that $A$ and $B_i$ share, and the change in negativity due to the measurement by $B_i$. This procedure is repeated by subsequent Bobs, until the negativity of the state shared between some $B_{i+1}$ and $A$, after an unsharp measurement by $B_i$, reduces to zero. 

 The change in the negativity of $\rho^{w_\alpha}_{AB_i}$, due to a ``valid'' and ``required'' measurement by $B_i$, can be evaluated to be
\begin{eqnarray}
 \Delta N_i(\lambda_i)&=&\frac{1+4N_i}{4}(1-f(\lambda_i)), \hspace{0.5 cm} N_{i+1}\neq 0; \nonumber \\
 &=& N_i, \hspace{2.8 cm} N_{i+1}=0.
 \label{delN2}
 \end{eqnarray}
   It can be easily checked that $\Delta N_i(\lambda_i)>0$ for $1/3>\lambda_i\geqslant 1$ for any $N_i\neq 0$. Therefore, subsequent measurements, to witness the shared entanglement results only into lowering of entanglement content. This is expected, as negativity is an entanglement monotone and therefore its value either decreases or remains the same under LOCC, as mentioned earlier. One can also observe that it is an increasing function of $\lambda_i$, and therefore, sharper the measurement, more is the decrease in the entanglement content.

Note that the unsharp measurement parameter, $\lambda_i$, for each $i$, should be equal to the threshold unsharpness parameter, $\lambda_i^{th}$, for the purpose of witnessing the shared entanglement sequentially in the optimal scenario (to obtain the maximum number of $B_i$ who can sequentially witness the shared entangement with A), discussed in the previous section.  The threshold unsharpness parameter further depends on the negativity of the shared state, $\rho^{w_\alpha}_{AB_i}$, via the relation
\begin{equation}
\lambda_i^{th}=\frac{1}{4N_i+1}.
\label{threslam_vs_neg}
\end{equation}
Therefore, in the optimal scenario, the change in negativity of state, $\rho^{w_\alpha}_{AB_i}$, with negativity, $N_i$, when $N_{i+1}\neq 0$, turns out to be 
\begin{equation}
\Delta N_i(\lambda_i^{th})=\frac{1}{8}\left[1+4N_i-\sqrt{N_i(1+N_i)}-\sqrt{3N_i(1+3N_i)}\right]. 
\label{delN3}
\end{equation}
 
 Note that given a state with negativity $N_i$, $\Delta N_i(\lambda_i^{th})$ is always positive, and a strictly increasing function of the discrete variable $i$ (for $N_{i+1} \neq 0$), which guarantees that $N_j=0$ can be reached at some finite number of $B_j$.

\section{Equistrength unsharp measurements}   
\label{baadal}
\begin{figure}[ht]
\includegraphics[width = 0.35\textwidth, angle=-90]{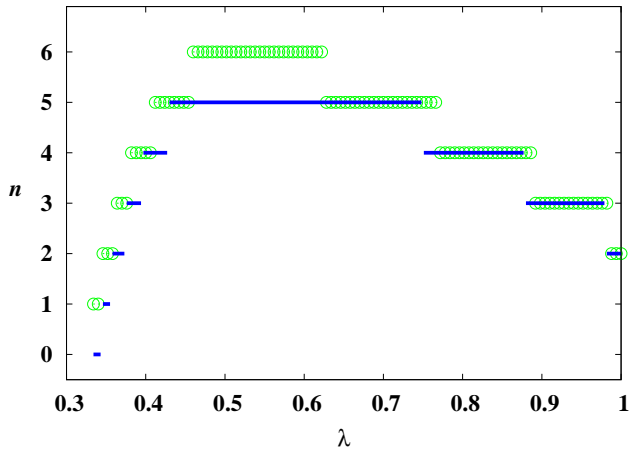} 
\caption{How weak can effective and equally strong Bobs be to maximize their number? We plot here the maximum number of observers, $n$, making unsharp measurements, against the common sharpness parameter, $\lambda$. The Bobs are required to do unsharp measurements of equal strength. The green circles represent situation for which the shared initial state has $E(\alpha)=1$, while the blue line is for initial states having $E(\alpha)=0.935$.
Both axes represent dimensionless quantities, while $E$ is in ebits.
}
\label{Fig.2}
\end{figure}
The optimal scenario where subsequent observers are allowed to measure with the threshold value of sharpness parameter can be experimentally challenging as well as costly. It can be challenging because, each subsequent Bob needs to tune the apparatus precisely to attain the maximum number of Bobs, and can be costly if they need to use separate apparatuses for the different sharpness parameter. In this section, therefore, we put some more restrictions on the Bobs. Specifically, independence of the sequential observers is lifted, to the extent that they are required to measure with equal sharpness parameter, $\lambda \in (\frac{1}{3},1]$.
In Fig. \ref{Fig.2}, the maximum number, $n$, of such sequential observers with the same sharpness parameter, $\lambda$, is plotted for fixed entanglement contents of the initial state, namely $E(\alpha)=0.935$ and 1 ebit. In the former case, the maximum $n$ over all parameter range of $\lambda$, denoted by say, $n_{max}$, is found to be five, whereas in the latter, the same maximum is six. Note that for any given initial entanglement, $n$ is the maximum number of Bobs measuring at any common sharpness parameter, $\lambda$, whereas $n_{max}$ denotes the maximum $n$ over all $\lambda$.
We can see that for intially shared nearly maximally entangled states, $n_{max}=6$. Again, this is better when compared to the same task in a device-dependent scenario, where a maximum of five observers can witness the entanglement with equal unsharp measurements for initially shared nearly maximally entangled states \cite{bera}. 

\begin{figure}[]
\includegraphics[width = 0.35\textwidth, angle=-90]{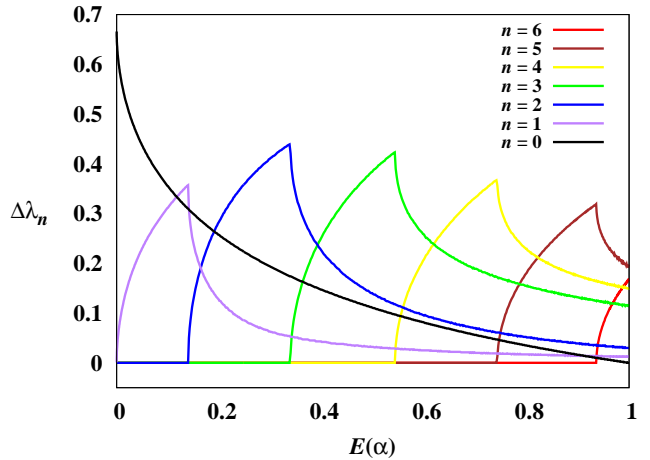} 
\caption{Variation of the range of the common sharpness parameter with respect to the initial entanglement in the equistrength unsharp measurements scenario. The range of common sharpness parameter, $\Delta\lambda_n$, is plotted against the initial entanglement, $E(\alpha)$, of the shared state with $n$ as the parameter. The ordinate is dimensionless while abscissa is in ebits.
}
\label{Fig.3}
\end{figure}

In the DD case, observers performing the sharpest measurement (measurement with sharpness parameter equal to 1) cause shared entanglement between two laboratories to vanish. That is, only the first Bob can detect the entanglement with Alice and the rest Bobs won't. On the other hand, here, we observe that even after the sharpest measurement, the shared entanglement will exist for some initial shared entangled states. This can be seen in Fig. \ref{Fig.2}, e.g. when the initial state possesses nearly maximal or maximal entanglement, two Bobs can detect entanglement with sharpness parameter being equal to unity. The fact that entanglement can be non-zero even after a sharp measurement suggests a better robustness of the MDI-EW compared to the device-dependent witness operator.

Note that for any initially shared pure entangled state, $n=n_{max}$ is reached for intermediate values (not very high and not very low) of sharpness measurement parameter, $\lambda$, i.e., $n=n_{max}$ is never achieved for values of $\lambda$ close to 1/3 or close to 1 (See Fig. \ref{Fig.2}).  Specifically, as one moves away from the intermediate values of $\lambda$ on the either side, i.e., either higher or lower values, maximum number of observers, $n$, measuring unsharply with equal strengths, either decrease or remain the same. This suggests that in order to achieve the maximum of $n$ over all the values of sharpness parameter, the observers should not set their sharpness parameter too high or too low. 
Such observation can be explained by the results reported in Sec. \ref{alternate}. If the first observer, $B_1$, sharing a state with $A$ having negativity $N_1$, measures unsharply with a parameter $\lambda_1$, then the state that is at the disposal of $A$ and $B_2$ surely possesses, on average, a lower value of negativity, $N_2$, compared to $N_1$, i.e., $N_2<N_1$. This can be seen from the relation given in  Eq. (\ref{delN2}). Since the negativity is decreasing with subsequent measurements, the new threshold parameter, $\lambda_2$, is greater than $\lambda_1$ (see Eq. (\ref{threslam_vs_neg})). Therefore, if $B_1$ fixes the sharpness parameter to be at the threshold value at which he can detect the shared entanglement, i.e, at $\lambda_1^{th}$, then only he can witness the entanglement while others cannot, as the threshold sharpness parameter to detect entanglement will be greater for subsequent Bobs. This explains the occurence of the least number of $B_i$ at lower values of sharpness parameter. On the other hand, if $B_1$ chooses to measure shaply, i.e, $\lambda_1 = 1$, then the state disturbed to the maximum possible, as discussed in Sec. \ref{alternate}, and therefore, a lower number of $B_i$s can only  witness entanglement sequentially with the same $\lambda$. 


Let us now study the dependence of the length or the range of common sharpness parameter, denoted by $\Delta\lambda_n$, on the initial entanglement, $E(\alpha)$, of the shared state, with $n$ being the parameter.
 For $n=n_{max}$, $\Delta\lambda_n$ decreases with decrease in $E(\alpha)$, and for the rest values of $n$,  $\Delta\lambda_n$ increases with decrease in $E(\alpha)$. See Fig. \ref{Fig.3}.

\section{Conclusion}
\label{tareef}
Entangled states have already been established as a resource in several quantum information processing tasks. Therefore, detection of entanglement in laboratory set-ups is an important task. If partial knowledge of an entangled state is available, employing entanglement witnesses for entanglement detection is, in principle, possible with trusted devices. On the other hand, violation of Bell inequality certifies entanglement in a device-independent way but at the cost that not all entangled states violate a Bell inequality. To bridge this gap, a measurement-device-independent entanglement witness (MDI-EW) has recently been introduced which yield a higher pay-off for every entangled state compared to separable states, by invoking a semi-quantum nonlocal game. Here we employed a MDI-EW to detect  entanglement in a novel entanglement distribution scenario where half of a pure entangled state is measured by a single observer, while the other half is measured by several observers sequentially and independently. We found that the number of observers who successfully detect entanglement with the other party, is larger than in   the similar sequential scenarios considered for violation of Bell inequality and for device-dependent entanglement witness operators. More interestingly, we observed that without employing unsharp measurements, one can still have detection of entanglement upto two observers which was not the case for the two other entanglement identification schemes. Both these results established that the MDI-EW method is more robust as compared to the other methods of entanglement detection. The sequential sharing of entanglement was studied both in the cases when all the observers are free to choose their optimal unsharp measurements, and when all of them are constrained to choose a specific unsharp measurement. Our results show that sequential sharing of quantum states in a measurement-device-independent way can be beneficial for quantum information processing tasks.

\end{document}